\documentclass[a4paper, 11pt]{article}
\usepackage{paracol}

\usepackage{xcolor}
\usepackage{setspace}
\usepackage{hanging}
\usepackage{changepage}
\usepackage{amsmath}
\allowdisplaybreaks
\usepackage{hyperref}
\definecolor{mycolor}{RGB}{0,88,204}
\hypersetup{
  colorlinks=true,
  linkcolor=black,
  urlcolor=mycolor,
  citecolor=mycolor
}
\usepackage{amsthm}
\usepackage{tikz}
\theoremstyle{definition}

\newcounter{mycounter}[section]








\usepackage{geometry}

\usepackage{dirtree}
\usepackage{fancyhdr}

\fancypagestyle{plain}{%
  \fancyhf{}
  \fancyhead[R]{\thepage}
  
}
\usepackage{mathptmx}
\DeclareMathAlphabet{\mathcal}{OMS}{cmsy}{m}{n}
\usepackage{titlesec}

\titleformat{\section}
  {\normalfont\large\centering}{\thesection.}{1em}{\MakeUppercase}
\titleformat{\subsection}
  {\normalfont\centering}{\thesubsection.}{1em}{\MakeUppercase}

  \titlespacing{\paragraph}{10pt}{0pt}{6pt}[0pt]
\usepackage{listings}
\usepackage[T1]{fontenc}
\usepackage[utf8]{inputenc}
\usepackage{amssymb}
\usepackage{chngcntr}
\usepackage{inconsolata}
\usepackage{upquote}
\definecolor{keywordcolor}{rgb}{0.7, 0.1, 0.1}   % red
\definecolor{tacticcolor}{rgb}{0.0, 0.1, 0.6}    % blue
\definecolor{commentcolor}{rgb}{0.4, 0.4, 0.4}   % grey
\definecolor{symbolcolor}{rgb}{0.0, 0.1, 0.6}    % blue
\definecolor{sortcolor}{rgb}{0.1, 0.5, 0.1}      % green
\definecolor{attributecolor}{rgb}{0.7, 0.1, 0.1} % red

\lstnewenvironment{code}[1][]%
{
   \noindent\newline
   \minipage{1\linewidth}
   \vspace{0.5\baselineskip}
   \lstset{
 	frame = lrtb,
 	rulecolor=\color{mycolor},
 	escapeinside={/*!}{!*/},
	language=lean,
	aboveskip = 5mm,
	belowskip = 5mm,
	xleftmargin=2mm,
	xrightmargin=2mm,
	}
	}
{\endminipage\newline}

\usepackage{float}
\usepackage{xpatch}
\usepackage{listings}
\usepackage{realboxes}
\definecolor{mycolorSubtle}{RGB}{245,250,255}
\DeclareRobustCommand{\myinline}{\lstinline}
\makeatletter
\xpretocmd\myinline{\Colorbox{mycolorSubtle}\bgroup\appto\lst@DeInit{\egroup}}{}{}
\makeatother

\usepackage[most]{tcolorbox}
\tcbuselibrary{breakable}
\usepackage{caption}
\usepackage{cleveref}

\definecolor{darkgreen}{RGB}{0, 100, 0}

\DeclareCaptionType{snippet}[Snippet][List of Snippets]

\crefformat{snippet}{\{\textcolor{darkgreen}{Input~#2#1#3}\}}
\Crefformat{snippet}{\textcolor{darkgreen}{\{#2#1#3\}}}

\newtcolorbox{startbox}{
	breakable, % Allows breaking over pages, which helps with long code blocks.
	enhanced jigsaw, % means that boxes split over pages do not have bottom & top lines.
    colback=gray!10,      % Light gray background
    colframe=darkgreen,     % Slightly darker border
    arc=2mm,              % Rounded corners
    boxrule=0.8pt,        % Border thickness
    left=2mm, right=2mm, top=2mm, bottom=2mm,
    boxsep=0mm,
    fonttitle=\bfseries
}

\lstnewenvironment{codeLong}[1][]%
{
    \lstset{
        frame=none,
        language=lean,
        aboveskip=0mm,
        belowskip=2mm,
        xleftmargin=1mm,
        basicstyle=\ttfamily\scriptsize,
        columns=fixed,
        fontadjust=true,
        breaklines=true,
        escapeinside={(*}{*)},
        #1
    }
}
{}

\NewDocumentEnvironment{inputbox}{m}
{
  \begin{startbox}
  \captionof{snippet}{}
  \label{#1}
  \vspace{-0.25cm}
  {\color{darkgreen}\hrule} \vspace{2mm}
}
{
  \end{startbox}
}
\NewDocumentEnvironment{inputboxFoot}{}
{
  {\color{darkgreen}\hrule} \vspace{2mm}
}
{
}
\newcommand{\physLeanLinks}[2][]{%
  \def\physLeanLinksTemp{#1}%
  \ifx\physLeanLinksTemp\empty
    \href{https://physlean.com/docs/find/?pattern=#2\#doc}{\texttt{#2}}%
  \else
    \href{https://physlean.com/docs/find/?pattern=#2\#doc}{\texttt{#1}}%
  \fi
}

\DeclareCaptionType{rsnippet}[Snippet][List of Snippets]

\makeatletter
\let\c@rsnippet\c@snippet

\makeatother
\crefformat{rsnippet}{[\textcolor{red}{Res.~#2#1#3}]}
\Crefformat{rsnippet}{\textcolor{red}{\{#2#1#3\}}}

\newtcolorbox{formbox}{
	breakable, % Allows breaking over pages, which helps with long code blocks.
    colback=gray!10,      % Light gray background
    colframe=red,     % Slightly darker border
    arc=2mm,              % Rounded corners
    boxrule=0.8pt,        % Border thickness
    left=2mm, right=2mm, top=2mm, bottom=2mm,
    boxsep=0mm,
    fonttitle=\bfseries
}

\NewDocumentEnvironment{derbox}{m}
{
  \begin{formbox}
  \captionof{rsnippet}{}
  \label{#1}
  \vspace{-0.25cm}
  {\color{red}\hrule} \vspace{2mm}
}
{
  \end{formbox}
}
\NewDocumentEnvironment{derboxFoot}{}
{
  {\color{red}\hrule} \vspace{2mm}
}
{
}
\newcommand{\codeLink}[1]{
  \vspace{-0.5cm}\hfill\href{https://github.com/HEPLean/HepLean/blob/1b951994ae3d882242b02d23957ef1ee7fa05f3d/HepLean/#1}{(source)}
  }
  
 \newcommand{\textLink}[1]{\href{https://github.com/HEPLean/HepLean/blob/1b951994ae3d882242b02d23957ef1ee7fa05f3d/HepLean/#1}{source}}
 \newcommand{\textLinkB}[1]{\href{https://github.com/HEPLean/HepLean/blob/1b951994ae3d882242b02d23957ef1ee7fa05f3d/HepLean/#1}{(source)}}
\title{Formalizing the stability of the two Higgs doublet model potential into~Lean:
  identifying an error in the literature}
\author{Joseph Tooby-Smith \\ \textit{Department of Computer Science, University of Bath,}
\\ \textit{Claverton Down, Bath, BA2 7AY, United Kingdom}}
\date{\today}
\usepackage[framemethod=TikZ]{mdframed}
\usepackage{hyperref}

\newmdenv[
  linecolor=black,
  linewidth=1pt,
  backgroundcolor=gray!2,
  roundcorner=5pt,
  innertopmargin=10pt,
  innerbottommargin=10pt,
]{thesisbox}
\begin{document}
\counterwithin{lstlisting}{section}
\maketitle
\vspace{-1cm}
\begin{abstract}
In 2006, using the best methods and techniques available at the time,
Maniatis, von Manteuffel, Nachtmann and Nagel published a now widely cited paper on
the stability of the two Higgs doublet model (2HDM) potential. Twenty years on,
it is now easier to apply the process of formalization into an interactive theorem prover to
 this work thanks to projects like Mathlib and Physlib (the latter formerly PhysLean and Lean-QuantumInfo),
 and to ask for a
 higher standard of mathematical correctness.
Doing so has revealed an error in the arguments of this 2006 paper, invalidating their main theorem
on the stability of the 2HDM potential. This case is noteworthy because
to the best of our knowledge it is the
first non-trivial error in a physics paper found through formalization.
It was one of the first papers where formalization was attempted, which raises
the uncomfortable question of how many physics papers would not pass this higher
level of scrutiny.
\end{abstract}
%\tableofcontents

%%%%%%%%%%%%%%%%%%%%%%%%%%%%%%%%%%
%%%%%%%%%%%%%%%%%%%%%%%%%%%%%%%%%%
%%%%%%%%%%%%%%%%%%%%%%%%%%%%%%%%%%
\section{Introduction}
%%%%%%%%%%%%%%%%%%%%%%%%%%%%%%%%%%
%%%%%%%%%%%%%%%%%%%%%%%%%%%%%%%%%%
%%%%%%%%%%%%%%%%%%%%%%%%%%%%%%%%%%
Through the use of a type of computer
programming language called an interactive theorem prover (ITP), over the
last few years it has become easier to guarantee the mathematical correctness of
results in the physics literature.

An ITP is a programming language
in which you can write definitions, theorems, and proofs from mathematics and physics
and in which mathematical correctness is guaranteed through a mathematical foundation of
type theory. Examples of ITPs include Rocq, Agda, Isabelle, and Lean. The process of
writing results into an ITP is often called formalization or sometimes digitization.

The largest physics library in the ITP Lean~\cite{lean} is the project Physlib~\cite{PhysLib} (recently renamed
from PhysLean~\cite{heplean} to merge with the Lean-QuantumInfo library~\cite{meiburg2024quantuminfo,Meiburg:2025mwn}), which
is open-source and community-run.
The focus of Physlib is not on axiomatizing physics (as one may expect), but rather
follows the approach taken by physicists when they write papers. A suitable
starting point is chosen and the logical conclusions are derived from that starting point.
Eventually that starting point can be moved back to more fundamental grounds, and
combined with other starting points to give a comprehensive interconnected library of results.
With this approach, it becomes possible to start formalizing research level results
from a broad range of areas in physics in a way which is generally useful
across the different branches of the discipline.

The origins of the paper you are reading lay in what was originally meant to be a tick-box exercise to
incorporate into Physlib standard results from the two Higgs doublet model (2HDM),
a popular extension of the Standard Model corresponding to increasing
the Higgs sector to two Higgs doublets. The 2006 paper~\cite{Maniatis:2006fs} by
Maniatis, von Manteuffel, Nachtmann and Nagel was chosen as a starting point
as it is used as a standard reference for the stability of the 2HDM potential, and
with the
mathematical nature of the work making the process of formalization easier and the
probability of correctness higher.

There were no signs in the literature (as far as we could tell) that
the main result of the paper~\cite{Maniatis:2006fs}
on the stability of the potential, Theorem 1, was incorrect, nor did we
have any indication of this before starting the process of formalization into Physlib.
It was a surprise, therefore, when formalization indicated that
there was an error in the argument of~\cite{Maniatis:2006fs} ultimately leading to Theorem 1 being incorrect,
and that we were able to formally verify the incorrectness of the argument.

This is significant for two reasons. Firstly, we believe this to be the first
time a non-trivial error in a research level physics paper has been identified through the process
of formal verification (minor imprecisions have been found before; see e.g.~\cite{Meiburg:2025mwn}). Secondly, this was one of the first research-level papers where formalization was attempted,
and it was not chosen with the intention of finding an error,
but rather because we thought the process of formalization would be easy and the likelihood of an error was low.
From this one could make the worrying extrapolation that there are many such errors in the physics literature.
It is also a strong motive
for making formal verification the gold standard for physics papers.

In this paper we will do three things.
Firstly, we will restate the correct arguments in~\cite{Maniatis:2006fs} around the stability of the
full 2HDM potential. These results
have all been formalized into Physlib and can be used for further analysis.
We will frame our discussion using the complexity of the
stability condition in terms of quantifier structure.

Secondly, we will take a small detour to give
a new and correct reduction in the complexity of the stability condition.
This again has been formalized into Physlib and can be used for further analysis.

Thirdly, we will then expose the error in~\cite{Maniatis:2006fs}.
The error is in a statement that says that a certain condition C is
necessary and sufficient for the stability of the 2HDM potential.
Throughout this paper we will use the term `stability' to exclusively refer to the
condition that the potential is bounded from below.
We will show that it is not sufficient by
showing it breaks down in what~\cite{Maniatis:2006fs} call the
marginally stable case by providing an explicit potential
which is not stable but which satisfies the condition C.
 This is
formally verified. Theorem 1 in~\cite{Maniatis:2006fs} is a reworking of this incorrect statement,
and as such is invalidated, in that it is demonstrably false. We will not
formally verify or discuss this reworking of condition C into Theorem 1 as
it can be easily seen from~\cite{Maniatis:2006fs}
and does not add anything to the story because of the error.

Theorem 1 in~\cite{Maniatis:2006fs} is about the stability of the full 2HDM potential, which
includes both the quadratic and quartic terms. Many authors only consider the quartic term when
discussing the stability of potentials, and the error in~\cite{Maniatis:2006fs} does
not invalidate any of the conclusions there. Because of this,
and the fact that the error occurs in those potentials which are on
the boundary between stable and unstable, while this error
is non-trivial in that it invalidates Theorem 1 in~\cite{Maniatis:2006fs}, it is not
expected that this error will have a large impact on downstream results; however,
we do not think this takes away from the significance of this paper.
It is also worth noting here that~\cite{Maniatis:2006fs} discusses
results beyond the stability of the potential, including
electroweak symmetry breaking and the mass spectrum of the model.
We do not discuss these in this paper, and we do not believe that they are
affected in any major way by the error.

We will split what follows
into two columns. In the left column, we will give the English or text argument, which can
be followed by those not interested in the formalization aspect of this paper.
Here, all relevant claims will be linked to code snippets which appear in the
right-hand column containing the formalized version of that claim. Those results marked as
e.g.~\cref{code:HiggsDoublet} constitute input data and define the starting point of the
formalization, whilst those results marked as e.g.~\cref{code:gram_matrix} are derived results.
Due to a difference in information density between the English and formalized versions,
the two columns will not be perfectly aligned; however, we think this approach is best
as it highlights the formalization without distracting from the English story.

It is worth noting that there was no extensive use of AI in the formalization
within this paper. Because of this, there is the extra guarantee that the results in this
paper are physically correct and correspond to the intended mathematical statements as
they have all been human-written and checked.

In the Appendix we give some tips on reading the Lean code, however if the reader
wants a fuller introduction, and a summary of the use of Lean in physics,
they are directed to~\cite{perspective}.

The remainder of this paper is split into the following sections. In \Cref{sec:Higgses}
we will introduce the Higgs doublet vectors and relate them to the Gram matrix.
In \Cref{sec:potential} we will introduce the 2HDM potential and give some results around it.
In \Cref{sec:stability} we will give the correct arguments in~\cite{Maniatis:2006fs} around
the stability of the full 2HDM potential. We will do this up to the point where we can
manifest the error, which we will discuss and formally prove in~\Cref{sec:TheError}.
\vspace{1cm}
\setlength{\columnsep}{0.4cm}
\begin{paracol}{2}

%%%%%%%%%%%%%%%%%%%%%%%%%%%%%%%%%%
\section{The Higgses} \label{sec:Higgses}
%%%%%%%%%%%%%%%%%%%%%%%%%%%%%%%%%%
\switchcolumn
\begin{center}\textbf{\ref{sec:Higgses}. The Higgses}\end{center}
\switchcolumn
The Higgs sector of the two Higgs doublet model consists of two
fields which are maps from spacetime to the complex vector space $\mathbb{C}^2$.
Here we are not interested in the spacetime dependence of these fields
and so we can work with the underlying vectors at a given spacetime point, which we will denote $\Phi_1$ and $\Phi_2$~\cref{code:HiggsDoublet}.
In this section we give some important parts of the `API' (lemmas and results) around these Higgs vectors.
\switchcolumn
%%%% Code snippet
\begin{inputbox}{code:HiggsDoublet}
\begin{codeLong}
abbrev (*\physLeanLinks[HiggsVec]{StandardModel.HiggsVec}*) :=
  EuclideanSpace ℂ (Fin 2)

structure (*\physLeanLinks[TwoHiggsDoublet]{TwoHiggsDoublet}*) where
  Φ1 : HiggsVec
  Φ2 : HiggsVec
\end{codeLong}
\begin{inputboxFoot}
The definition of Higgs vector space corresponding to $\mathbb{C}^2$, and the
two Higgs doublet corresponding to a pair of Higgs vectors.
\end{inputboxFoot}
\end{inputbox}
\switchcolumn

With the vectors $\Phi_1$ and $\Phi_2$, we can define the Gram matrix $G$ as follows (denoted $\underline{K}$ in~\cite{Maniatis:2006fs})~\cref{code:gram_matrix}
\begin{align}
G := \begin{pmatrix} \|\Phi_1\|^2 & \Phi_2^\dagger \Phi_1\\ \Phi_1^\dagger \Phi_2 & \|\Phi_2\|^2 \end{pmatrix}.
\end{align}
%% Start of code block
\switchcolumn
\begin{derbox}{code:gram_matrix}
\begin{codeLong}
def (*\physLeanLinks[gramMatrix]{TwoHiggsDoublet.gramMatrix}*) (H : TwoHiggsDoublet) :
    Matrix (Fin 2) (Fin 2) ℂ :=
  !![⟪H.Φ1, H.Φ1⟫_ℂ, ⟪H.Φ2, H.Φ1⟫_ℂ;
    ⟪H.Φ1, H.Φ2⟫_ℂ, ⟪H.Φ2, H.Φ2⟫_ℂ]
\end{codeLong}
\begin{derboxFoot}
The Gram matrix of two Higgs doublet.
\end{derboxFoot}
\end{derbox}
\switchcolumn

The Gram matrix is self-adjoint~\cref{code:gramMatrix_selfAdjoint}. This allows it to
be written in terms of the Pauli-matrices and allows us to define a 4-component vector $K_\mu$ such that
$G = \frac{1}{2} K_\mu \sigma^\mu$ where $\sigma^0 = 1$~\cref{code:gramVector}.
We call the vector $K$ the `Gram vector'.
Explicitly its components are~\cref{code:gramVector_compnents}
\begin{align}\label{eq:k_def}
K_0 &=  \|\Phi_1\|^2 + \|\Phi_2\|^2, \nonumber\\
K_1 &= 2 * \mathrm{Re}(\Phi_1^\dagger \Phi_2), \nonumber\\
K_2 &= 2 * \mathrm{Im}(\Phi_1^\dagger \Phi_2), \nonumber\\
K_3 &= \|\Phi_1\|^2 - \|\Phi_2\|^2.
\end{align}
The Gram vector has the properties that $ 0 \leq K_0$~\cref{code:gramVector_inl_nonneg} and $\sum_{a =1}^3 K_a^2 \leq K_0^2$~\cref{code:gramVector_inr_sum_sq_le_inl}.
\switchcolumn
%%% Begin code block
\begin{derbox}{code:gramMatrix_selfAdjoint}
\begin{codeLong}
lemma (*\physLeanLinks[gramMatrix\_selfAdjoint]{TwoHiggsDoublet.gramMatrix\_selfAdjoint}*)
	  (H : TwoHiggsDoublet) :
	IsSelfAdjoint (gramMatrix H) := by ...
\end{codeLong}
\begin{derboxFoot}
The lemma that the Gram matrix is self-adjoint.
\end{derboxFoot}
\end{derbox}
%%% end code block
%%% Begin code block
\begin{derbox}{code:gramVector}
\begin{codeLong}
def (*\physLeanLinks[gramVector]{TwoHiggsDoublet.gramVector}*) (H : TwoHiggsDoublet) :
    Fin 1 ⊕ Fin 3 → ℝ := fun μ =>
  2 * PauliMatrix.pauliBasis.repr ⟨gramMatrix H, gramMatrix_selfAdjoint H⟩ μ
\end{codeLong}
\begin{derboxFoot}
The definition of the Gram vector.
\end{derboxFoot}
\end{derbox}
%%% end code block
%%% Begin code block
\begin{derbox}{code:gramVector_compnents}
\begin{codeLong}
lemma (*\physLeanLinks[gramVector\_inl\_zero\_eq]{TwoHiggsDoublet.gramVector\_inl\_zero\_eq}*)
    (H : TwoHiggsDoublet) :
  H.gramVector (Sum.inl 0) =
  ‖H.Φ1‖ ^ 2 + ‖H.Φ2‖ ^ 2 := by...

lemma (*\physLeanLinks[gramVector\_inr\_zero\_eq]{TwoHiggsDoublet.gramVector\_inr\_zero\_eq}*)
    (H : TwoHiggsDoublet) :
  H.gramVector (Sum.inr 0) =
  2 * (⟪H.Φ1, H.Φ2⟫_ℂ).re := by ...

lemma (*\physLeanLinks[gramVector\_inr\_one\_eq]{TwoHiggsDoublet.gramVector\_inr\_one\_eq}*)
    (H : TwoHiggsDoublet) :
  H.gramVector (Sum.inr 1) =
  2 * (⟪H.Φ1, H.Φ2⟫_ℂ).im := by ...

lemma (*\physLeanLinks[gramVector\_inr\_two\_eq]{TwoHiggsDoublet.gramVector\_inr\_two\_eq}*)
    (H : TwoHiggsDoublet) :
  H.gramVector (Sum.inr 2) =
  ‖H.Φ1‖ ^ 2 - ‖H.Φ2‖ ^ 2 := by...
\end{codeLong}
\begin{derboxFoot}
Lemmas giving the components of the gram vectors.
\end{derboxFoot}
\end{derbox}
%%% end code block
%%% Begin code block
\begin{derbox}{code:gramVector_inl_nonneg}
\begin{codeLong}
lemma (*\physLeanLinks[gramVector\_inl\_nonneg]{TwoHiggsDoublet.gramVector\_inl\_nonneg}*) (H : TwoHiggsDoublet) :
  0 ≤ H.gramVector (Sum.inl 0) := by...
\end{codeLong}
\begin{derboxFoot}
The first component of the gram vector is non-negative.
\end{derboxFoot}
\end{derbox}
%%% end code block
%%% Begin code block
\begin{derbox}{code:gramVector_inr_sum_sq_le_inl}
\begin{codeLong}
lemma (*\physLeanLinks[gramVector\_inr\_sum\_sq\_le\_inl]{TwoHiggsDoublet.gramVector\_inr\_sum\_sq\_le\_inl}*)
    (H : TwoHiggsDoublet) :
  ∑ μ : Fin 3, H.gramVector (Sum.inr μ) ^ 2
  ≤ H.gramVector (Sum.inl 0) ^ 2 := by...
\end{codeLong}
\begin{derboxFoot}
The sum of squares of the last three components of the gram vector is less than or equal to the square of the first component.
\end{derboxFoot}
\end{derbox}
%%% end code block
\switchcolumn

It turns out
that for every vector $K$ satisfying the properties that $ 0 \leq K_0$ and $\sum_{a =1}^3 K_a^2 \leq K_0^2$,
there exists a pair of Higgs doublet vectors $\Phi_1, \Phi_2$ such that $K$ is given by the Equation~\ref{eq:k_def}~\cref{code:gramVector_surjective}.
Furthermore, two pairs of Higgs doublet vectors $\Phi_1, \Phi_2$ and $\Phi_1^\prime, \Phi_2^\prime$
give the same vector $K$ if and only if they are in the same
group orbit under the Standard Model global gauge group~\cref{code:mem_orbit_gaugeGroupI_iff_gramVector}.
\switchcolumn
%%% Begin code block
\begin{derbox}{code:gramVector_surjective}
\begin{codeLong}
lemma (*\physLeanLinks[gramVector\_surjective]{TwoHiggsDoublet.gramVector\_surjective}*)
    (v : Fin 1 ⊕ Fin 3 → ℝ)
    (h_inl : 0 ≤ v (Sum.inl 0))
    (h_det : ∑ μ : Fin 3, v (Sum.inr μ) ^ 2 ≤
    v (Sum.inl 0) ^ 2) :
  ∃ H : TwoHiggsDoublet, H.gramVector = v := by ...
\end{codeLong}
\begin{derboxFoot}
Creation of the gram vector surjects onto the set of vectors satisfying certain properties.
\end{derboxFoot}
\end{derbox}
%%% end code block
%%% Begin code block
\begin{derbox}{code:mem_orbit_gaugeGroupI_iff_gramVector}
\begin{codeLong}
lemma (*\physLeanLinks[mem\_orbit\_gaugeGroupI\_iff\_gramVector]{TwoHiggsDoublet.mem\_orbit\_gaugeGroupI\_iff\_gramVector}*)
    (H1 H2 : TwoHiggsDoublet) :
  H1 ∈ MulAction.orbit GaugeGroupI H2 ↔ H1.gramVector = H2.gramVector := by...
\end{codeLong}
\begin{derboxFoot}
Gram vectors characterize the orbit under the gauge group of two Higgs doublets. We
do not go into the details of the gauge group here.
\end{derboxFoot}
\end{derbox}
%%% end code block
\switchcolumn
%%%%%%%%%%%%%%%%%%%%%%%%%%%%%%%%%%
\section{The 2HDM Potential} \label{sec:potential}
%%%%%%%%%%%%%%%%%%%%%%%%%%%%%%%%%%
\switchcolumn
\begin{center}\textbf{\ref{sec:potential}. The 2HDM Potential}\end{center}
\switchcolumn
We will now give general properties and results around the 2HDM potential.

A common way (see e.g.~\cite{Draper:2016cag}) to parameterize the 2HDM potential is by the six real parameters $m_{11}^2$, $m_{22}^2$,
$\lambda_1$, $\lambda_2$, $\lambda_3$, $\lambda_4$,
and the four complex parameters $m_{12}^2$, $\lambda_5$, $\lambda_6$, $\lambda_7$~\cref{code:parameters}.
For $\Phi_1$ and $\Phi_2$ two Higgs doublet vectors the potential, $V$, corresponding to these parameters
is given by~\cref{code:potential}~\cite{Draper:2016cag}
\begin{align}
V_2 :=&\, m_{11}^2 \|\Phi_1\| ^ 2 + m_{22}^2 \|\Phi_2\| ^ 2 - [m_{12}^2 \Phi_1^\dagger \Phi_2 + \mathrm{h.c.}],
\nonumber\\V_4 :=& \frac{1}{2} \lambda_1 \|\Phi_1\| ^ 4 + \frac{1}{2} \lambda_2 \|\Phi_2\| ^ 4 + \lambda_3 \|\Phi_1\| ^ 2 \|\Phi_2\| ^ 2
\nonumber\\& + \lambda_4 (\Phi_1^\dagger \Phi_2)(\Phi_2^\dagger \Phi_1) + [\frac{1}{2} \lambda_5 (\Phi_1^\dagger \Phi_2)^2
\nonumber\\&+ [\lambda_6 \|\Phi_1\| ^ 2 +\lambda_7 \|\Phi_2\| ^ 2]\Phi_1^\dagger \Phi_2 + \mathrm{h.c.}],
\nonumber\\V :=& V_2 + V_4.
\end{align}
where $V_2$ is the mass term and $V_4$ is the quartic term.
\switchcolumn

%%%% Code snippet
\begin{inputbox}{code:parameters}
\begin{codeLong}
structure (*\physLeanLinks[PotentialParameters]{TwoHiggsDoublet.PotentialParameters}*) where
  m₁₁2 : ℝ
  m₂₂2 : ℝ
  m₁₂2 : ℂ
  𝓵₁ : ℝ
  𝓵₂ : ℝ
  𝓵₃ : ℝ
  𝓵₄ : ℝ
  𝓵₅ : ℂ
  𝓵₆ : ℂ
  𝓵₇ : ℂ
\end{codeLong}
\begin{inputboxFoot}
{The parameters of the 2HDM potential. We use
\myinline|𝓵| instead of \myinline|λ|, as the latter
has special meaning in Lean.}
\end{inputboxFoot}
\end{inputbox}
%%%% End of code snippet

%%%% Code snippet
\begin{inputbox}{code:potential}
\begin{codeLong}
def (*\physLeanLinks[massTerm]{TwoHiggsDoublet.massTerm}*) (P : PotentialParameters)
    (H : TwoHiggsDoublet) : ℝ :=
  P.m₁₁2 * ‖H.Φ1‖ ^ 2 + P.m₂₂2 * ‖H.Φ2‖ ^ 2 - (P.m₁₂2 * ⟪H.Φ1, H.Φ2⟫_ℂ
  + conj P.m₁₂2 * ⟪H.Φ2, H.Φ1⟫_ℂ).re

def (*\physLeanLinks[quarticTerm]{TwoHiggsDoublet.quarticTerm}*) (P : PotentialParameters)
    (H : TwoHiggsDoublet) : ℝ :=
  1/2 * P.𝓵₁ * ‖H.Φ1‖ ^ 2 * ‖H.Φ1‖ ^ 2
  + 1/2 * P.𝓵₂ * ‖H.Φ2‖ ^ 2 * ‖H.Φ2‖ ^ 2
  + P.𝓵₃ * ‖H.Φ1‖ ^ 2 * ‖H.Φ2‖ ^ 2
  + P.𝓵₄ * ‖⟪H.Φ1, H.Φ2⟫_ℂ‖ ^ 2
  + (1/2 * P.𝓵₅ * ⟪H.Φ1, H.Φ2⟫_ℂ ^ 2
  + 1/2 * conj P.𝓵₅ * ⟪H.Φ2, H.Φ1⟫_ℂ ^ 2).re
  + (P.𝓵₆ * ‖H.Φ1‖ ^ 2 * ⟪H.Φ1, H.Φ2⟫_ℂ
  + conj P.𝓵₆ * ‖H.Φ1‖ ^ 2 * ⟪H.Φ2, H.Φ1⟫_ℂ).re
  + (P.𝓵₇ * ‖H.Φ2‖ ^ 2 * ⟪H.Φ1, H.Φ2⟫_ℂ
  + conj P.𝓵₇ * ‖H.Φ2‖ ^ 2 * ⟪H.Φ2, H.Φ1⟫_ℂ).re

def (*\physLeanLinks[potential]{TwoHiggsDoublet.potential}*) (P : PotentialParameters)
    (H : TwoHiggsDoublet) : ℝ :=
  massTerm P H + quarticTerm P H
\end{codeLong}
\begin{inputboxFoot}
The definition of the mass term, the quartic term, and the potential of the 2HDM
in terms of the parameters of the potential and the two Higgs doublet.
\end{inputboxFoot}
\end{inputbox}
%%%% End of code snippet
\switchcolumn

Because the Gram vector $K$ describes the orbits of the 2HDM Higgs vectors under the gauge group,
it is no surprise, therefore, that the potential can be written in terms of it. To do this,
it is convenient to reparameterize the potential in terms of the 4-component
vector $\xi$ and the $4\times 4$ matrix $\eta$~\cite{Maniatis:2006fs}. The vector $\xi$ is
defined as~\cref{code:xi_def}
\begin{align}
\xi_0 &:= (m_{11}^2 + m_{22}^2)/2 \nonumber\\
\xi_1 &:= -\mathrm{Re}(m_{12}^2) \nonumber\\
\xi_2 &:= \mathrm{Im}(m_{12}^2) \nonumber\\
\xi_3 &:= (m_{11}^2 - m_{22}^2)/2
\end{align}
\switchcolumn
\begin{derbox}{code:xi_def}
\begin{codeLong}
def (*\physLeanLinks[$\xi$]{TwoHiggsDoublet.PotentialParameters.ξ}*) (P : PotentialParameters) :
    Fin 1 ⊕ Fin 3 → ℝ := fun μ =>
  match μ with
  | .inl 0 => (P.m₁₁2 + P.m₂₂2) / 2
  | .inr 0 => -Complex.re P.m₁₂2
  | .inr 1 => Complex.im P.m₁₂2
  | .inr 2 => (P.m₁₁2 - P.m₂₂2) / 2
\end{codeLong}
\begin{derboxFoot}
The definition of the vector $\xi$ in terms of the parameters of the 2HDM potential. We index
here by \myinline|Sum.inl 0|, \myinline|Sum.inr 0|, \myinline|Sum.inr 1| and
\myinline|Sum.inr 2| instead of $0, 1, 2$ and $3$.
\end{derboxFoot}
\end{derbox}
\switchcolumn

The $4\times 4$ matrix
$\eta$ is defined with components~\cref{code:eta_def}
\begin{align}
\eta_{00} &:= \frac{1}{8}(\lambda_1 + \lambda_2 + 2 \lambda_3), \nonumber \\
\eta_{01}, \eta_{10} &:= \frac{1}{4}(\mathrm{Re} \lambda_6 + \mathrm{Re} \lambda_7),\nonumber\\
\eta_{02}, \eta_{20} &:= \frac{1}{4}(-\mathrm{Im} \lambda_6 - \mathrm{Im} \lambda_7),\nonumber \\
\eta_{03}, \eta_{30} &:= \frac{1}{8}(\lambda_1 - \lambda_2), \nonumber \\
\eta_{11} &:= \frac{1}{4}(\lambda_4+\mathrm{Re} \lambda_5), \nonumber\\
\eta_{12}, \eta_{21} &:= -\frac{1}{4}\mathrm{Im}\lambda_5, \nonumber\\
\eta_{13}, \eta_{31} &:= \frac{1}{4}(\mathrm{Re} \lambda_6 - \mathrm{Re} \lambda_7), \nonumber\\
\eta_{22} &:= \frac{1}{4}(\lambda_4 - \mathrm{Re} \lambda_5), \nonumber\\
\eta_{23}, \eta_{32} &:= \frac{1}{4}(\mathrm{Im} \lambda_7 - \mathrm{Im} \lambda_6),\nonumber \\
\eta_{33} &:= \frac{1}{8}(\lambda_1 + \lambda_2 - 2 \lambda_3),
\end{align}
This matrix is symmetric~\cref{code:symm}.
\switchcolumn
\begin{derbox}{code:eta_def}
\begin{codeLong}
def (*\physLeanLinks[$\eta$]{TwoHiggsDoublet.PotentialParameters.η}*) (P : PotentialParameters) :
    Fin 1 ⊕ Fin 3 → Fin 1 ⊕ Fin 3 → ℝ
  | .inl 0, .inl 0 => (P.𝓵₁ + P.𝓵₂ + 2 * P.𝓵₃) / 8
  | .inl 0, .inr 0 => (P.𝓵₆.re + P.𝓵₇.re) / 4
  | .inl 0, .inr 1 => - (P.𝓵₆.im + P.𝓵₇.im) / 4
  | .inl 0, .inr 2 => (P.𝓵₁ - P.𝓵₂) / 8
  | .inr 0, .inl 0 => (P.𝓵₆.re + P.𝓵₇.re) / 4
  | .inr 1, .inl 0 => -(P.𝓵₆.im + P.𝓵₇.im) / 4
  | .inr 2, .inl 0 => (P.𝓵₁ - P.𝓵₂) / 8
  | .inr 0, .inr 0 => (P.𝓵₅.re + P.𝓵₄) / 4
  | .inr 1, .inr 1 => (P.𝓵₄ - P.𝓵₅.re) / 4
  | .inr 2, .inr 2 => (P.𝓵₁ + P.𝓵₂ - 2 * P.𝓵₃) / 8
  | .inr 0, .inr 1 => - P.𝓵₅.im / 4
  | .inr 2, .inr 0 => (P.𝓵₆.re - P.𝓵₇.re) / 4
  | .inr 2, .inr 1 => (P.𝓵₇.im - P.𝓵₆.im) / 4
  | .inr 1, .inr 0 => - P.𝓵₅.im / 4
  | .inr 0, .inr 2 => (P.𝓵₆.re - P.𝓵₇.re) / 4
  | .inr 1, .inr 2 => (P.𝓵₇.im - P.𝓵₆.im) / 4
\end{codeLong}
\begin{derboxFoot}
The definition of the matrix $\eta$ in terms of the parameters of the 2HDM potential.
\end{derboxFoot}
\end{derbox}

\begin{derbox}{code:symm}
\begin{codeLong}
lemma (*\physLeanLinks[$\eta$\_symm]{TwoHiggsDoublet.PotentialParameters.η\_symm}*) (P : PotentialParameters)
    (μ ν : Fin 1 ⊕ Fin 3) : P.η μ ν = P.η ν μ := by
  fin_cases μ <;> fin_cases ν <;> simp [η]
\end{codeLong}
\begin{derboxFoot}
The lemma that the matrix $\eta$ is symmetric.
\end{derboxFoot}
\end{derbox}
\switchcolumn

In terms of the vector $\xi$ and the matrix $\eta$, the potential as a function
of the Gram vector $K$ can be written as~\cref{code:potential_eq_gramVector}
\begin{equation}
V = \sum_{\mu = 0}^3 \xi_\mu K_\mu + \sum_{\mu, \nu = 0}^3 \eta_{\mu \nu} K_\mu K_\nu.
\end{equation}
\switchcolumn
%%% Begin code block
\begin{derbox}{code:potential_eq_gramVector}
\begin{codeLong}
lemma (*\physLeanLinks[potential\_eq\_gramVector]{TwoHiggsDoublet.potential\_eq\_gramVector}*)
    (P : PotentialParameters)
    (H : TwoHiggsDoublet) :
    potential P H = ∑ μ, P.ξ μ * H.gramVector μ +
    ∑ a, ∑ b, H.gramVector a * H.gramVector b * P.η a b := by ...
\end{codeLong}
\begin{derboxFoot}
The potential in terms of the Gram vector.
\end{derboxFoot}
\end{derbox}
%%% end code block
\switchcolumn

It is with this form of the potential that the analysis of the stability of the full 2HDM potential
began in~\cite{Maniatis:2006fs}.
The relation of this form of the potential to Lorentz vectors is given in~\cite{Ivanov:2006yq,Ivanov:2007de}.

%%%%%%%%%%%%%%%%%%%%%%%%%
\section{Stability of the 2HDM Potential} \label{sec:stability}
%%%%%%%%%%%%%%%%%%%%%%%%%
\switchcolumn
\begin{center}\textbf{\ref{sec:stability}. Stability of the 2HDM Potential}\end{center}
\switchcolumn

We now turn to the question of the stability of the full 2HDM potential.
We will give the correct statements about the stability of the full 2HDM potential
in~\cite{Maniatis:2006fs}. These will be used in the next section to recount the error in~\cite{Maniatis:2006fs}.
We will also demonstrate a new result related to the stability of the 2HDM potential,
which gives the best known complexity reduction of the stability condition.

The condition that the potential is stable (i.e. bounded from below) is that there exists a
real number $c$ such that for all Higgs doublets $\Phi_1, \Phi_2$ we have~\cref{code:stability}
\begin{equation}
c \leq V(\Phi_1, \Phi_2).
\end{equation}
\switchcolumn
\begin{inputbox}{code:stability}
\begin{codeLong}
def (*\physLeanLinks[PotentialIsStable]{TwoHiggsDoublet.PotentialIsStable}*) (P : PotentialParameters) :
    Prop := ∃ c : ℝ, ∀ H : TwoHiggsDoublet,
  c ≤ potential P H
\end{codeLong}
\begin{inputboxFoot}
The stability condition for the 2HDM potential.
\end{inputboxFoot}
\end{inputbox}
\switchcolumn

To understand results related to the stability of the 2HDM potential,
it is useful to borrow some concepts from the arithmetical hierarchy.
In particular the above stability condition has the complexity
\begin{align}
	\exists (c : \mathbb{R}), &\forall (\Phi_{11} \in \mathbb{C}), \forall (\Phi_{12} \in \mathbb{C}), \nonumber\\ & \forall (\Phi_{21} \in \mathbb{C}), \forall (\Phi_{22} \in \mathbb{C}), \_
\end{align}
We are interested in arguments which reduce the complexity of this condition.

Our first reduction is to state the stability condition in terms of the Gram vector $K$
instead of the Higgs doublet vectors $\Phi_1$ and $\Phi_2$. To do this
we use the potential written in terms of the Gram vector, and the fact that
every vector $K$ satisfying the properties that $ 0 \leq K_0$ and $\sum_{a =1}^3 K_a^2 \leq K_0^2$
can be reached by some choice of Higgs doublets. From this we have that the stability condition is
equivalent to the condition that there exists a real number $c$ such that for all vectors
$K$ satisfying the properties that $ 0 \leq K_0$ and $\sum_{a =1}^3 K_a^2 \leq K_0^2$ we have~\cref{code:potentialIsStable_iff_forall_gramVector}
\begin{equation}
c \leq \sum_{\mu = 0}^3 \xi_\mu K_\mu + \sum_{\mu, \nu = 0}^3 \eta_{\mu \nu} K_\mu K_\nu.
\end{equation}
This is a reduction in the complexity of the stability condition to
\begin{align}
	\exists (c : \mathbb{R}), &\forall (K_0 \in \mathbb{R}), \forall (K_1 \in \mathbb{R}), \nonumber\\ & \forall (K_2 \in \mathbb{R}), \forall (K_3 \in \mathbb{R}), \_
\end{align}
noting that previously our $\forall$ quantifiers were over complex numbers, and now they are over real numbers.
\switchcolumn
%%% Begin code block
\begin{derbox}{code:potentialIsStable_iff_forall_gramVector}
\begin{codeLong}
lemma (*\physLeanLinks[potentialIsStable\_iff\_forall\_gramVector]{TwoHiggsDoublet.potentialIsStable\_iff\_forall\_gramVector}*)
    (P : PotentialParameters) :
    PotentialIsStable P ↔ ∃ c : ℝ,
    ∀ K : Fin 1 ⊕ Fin 3 → ℝ, 0 ≤ K (Sum.inl 0) →
    ∑ μ : Fin 3, K (Sum.inr μ) ^ 2 ≤
      K (Sum.inl 0) ^ 2 →
      c ≤ ∑ μ, P.ξ μ * K μ
        + ∑ a, ∑ b, K a * K b * P.η a b := by ...
\end{codeLong}
\begin{derboxFoot}
Stability condition in terms of the Gram vector.
\end{derboxFoot}
\end{derbox}
%%% end code block
\switchcolumn

We now keep the complexity of the stability condition the same,
but we modify the condition.
To do this we define two functions $J_2$ and $J_4$ of 3-vectors.
The first function $J_2$ is based on the quadratic term of the potential and is given by~\cref{code:massTermReduced}
\begin{equation}
J_2(\vec k) = \xi_0 + \sum_{a = 1}^3 \xi_a k_a.
\end{equation}
The second function $J_4$ is based on the quartic term of the potential and is given by~\cref{code:quarticTermReduced}
\begin{equation}
J_4(\vec k) = \eta_{00} + 2 \sum_{a = 1}^3 \eta_{0a} k_a + \sum_{a,b = 1}^3 \eta_{ab} k_a k_b.
\end{equation}
Note that in the stability condition we can freely require $c$ to be non-positive (as we can take it to
be as negative as we like).
Then note that if $K_0 = 0$, then $K = 0$ and $c \leq V = 0$,
so in the stability condition we can restrict to the case when $0 < K_0$.
In this case we can define the normalized vector $\vec k := \frac{1}{K_0} (K_1, K_2, K_3)$,
which satisfies $\|\vec k \| ^ 2 \leq 1$. The stability condition then becomes~\cref{code:potentialIsStable_iff_exists_forall_forall_reduced}
\begin{equation}\label{eq:K0_J2_J4}
c \leq K_0 J_2(\vec k) + K_0^2 J_4(\vec k)
\end{equation}
for all $\|\vec k \| ^ 2 \leq 1$ and $0 < K_0$.
The complexity of this condition is the same as above, though now with different variables
\begin{align}\label{eq:K0_k_complexity}
	\exists (c : \mathbb{R}), &\forall (K_0 \in \mathbb{R}), \forall (k_1 \in \mathbb{R}), \nonumber\\ & \forall (k_2 \in \mathbb{R}), \forall (k_3 \in \mathbb{R}), \_
\end{align}
\switchcolumn
%%% Begin code block
\begin{derbox}{code:massTermReduced}
\begin{codeLong}
def (*\physLeanLinks[massTermReduced]{TwoHiggsDoublet.massTermReduced}*) (P : PotentialParameters) (k : EuclideanSpace ℝ (Fin 3)) : ℝ :=
  P.ξ (Sum.inl 0) + ∑ μ, P.ξ (Sum.inr μ) * k μ
\end{codeLong}
\begin{derboxFoot}
The definition of the function $J_2$.
\end{derboxFoot}
\end{derbox}
%%% end code block
%%% Begin code block
\begin{derbox}{code:quarticTermReduced}
\begin{codeLong}
def (*\physLeanLinks[quarticTermReduced]{TwoHiggsDoublet.quarticTermReduced}*) (P : PotentialParameters)
    (k : EuclideanSpace ℝ (Fin 3)) : ℝ :=
  P.η (Sum.inl 0) (Sum.inl 0)
  + 2 * ∑ b, k b * P.η (Sum.inl 0) (Sum.inr b)
  + ∑ a, ∑ b, k a * k b *
    P.η (Sum.inr a) (Sum.inr b)
\end{codeLong}
\begin{derboxFoot}
The definition of the function $J_4$.
\end{derboxFoot}
\end{derbox}
%%% end code block
\switchcolumn

\switchcolumn
%%% Begin code block
\begin{derbox}{code:potentialIsStable_iff_exists_forall_forall_reduced}
\begin{codeLong}
lemma (*\physLeanLinks[potentialIsStable\_iff\_exists\_forall\_forall\_reduced]{TwoHiggsDoublet.potentialIsStable\_iff\_exists\_forall\_forall\_reduced}*)
    (P : PotentialParameters) :
    PotentialIsStable P ↔ ∃ c ≤ 0, ∀ K0 : ℝ,
    ∀ k : EuclideanSpace ℝ (Fin 3), 0 < K0 →
    ‖k‖ ^ 2 ≤ 1 → c ≤ K0 * massTermReduced P k
      + K0 ^ 2 * quarticTermReduced P k := by ...
\end{codeLong}
\begin{derboxFoot}
The stability of the potential in terms of the functions $J_2$ and $J_4$.
\end{derboxFoot}
\end{derbox}
%%% end code block
\switchcolumn

We are now in a position to understand the claim of~\cite{Maniatis:2006fs},
which is incorrect. However, before we muddy the waters, let us give one reduction in the complexity
of the stability condition which is formalized to be correct, and which is
new in this paper.

The stability condition is equivalent to the condition that there exists a non-negative real number $c$ such that for all vectors $\vec k$ satisfying $\|\vec k \|^ 2\leq 1$ we have~\cref{code:potentialIsStable_iff_massTermReduced_sq_le_quarticTermReduced}
\begin{align} \label{eq:new_stability_condition}
0 \leq J_4(\vec k) &\text{ and } \nonumber \\ (J_2(\vec k) < 0 &\text{ implies }  J_2(\vec k) ^ 2 \leq 4 c J_4(\vec k)).
\end{align}
\sloppy The proof of this is via completing the square of $K_0 J_2(\vec k) + K_0^2 J_4(\vec k)$ with respect to $K_0$, and then
setting $K_0 = - J_2(\vec k)/(2J_4(\vec k))$ when $J_2(\vec k) < 0$.
\switchcolumn
%%% Begin code block
\begin{derbox}{code:potentialIsStable_iff_massTermReduced_sq_le_quarticTermReduced}
\begin{codeLong}
lemma (*\physLeanLinks[potentialIsStable\_iff\_massTermReduced\_sq\_le]{TwoHiggsDoublet.potentialIsStable\_iff\_massTermReduced\_sq\_le\_quarticTermReduced}*)
(*\physLeanLinks[\_quarticTermReduced]{TwoHiggsDoublet.potentialIsStable\_iff\_massTermReduced\_sq\_le\_quarticTermReduced}*) (P : PotentialParameters) :
    PotentialIsStable P ↔ ∃ c, 0 ≤ c ∧
    ∀ k : EuclideanSpace ℝ (Fin 3), ‖k‖ ^ 2 ≤ 1 →
      0 ≤ quarticTermReduced P k ∧
      (massTermReduced P k < 0 →
      massTermReduced P k ^ 2 ≤
      4 * quarticTermReduced P k * c) := by...
\end{codeLong}
\begin{derboxFoot}
A reduced complexity condition for the stability of the potential.
\end{derboxFoot}
\end{derbox}
%%% end code block
\switchcolumn
This reduces the complexity of the stability condition to
\begin{align}
	\exists (c : \mathbb{R}), & \forall (k_1 \in \mathbb{R}), \nonumber\\ & \forall (k_2 \in \mathbb{R}), \forall (k_3 \in \mathbb{R}), \_
\end{align}
We believe that it is an open question as to whether the complexity can be reduced further, and we encourage
any reader attempting this to contribute their results to Physlib.

%%%%%%%%%%%%%%%%%
\section{The Error}\label{sec:TheError}
%%%%%%%%%%%%%%%%%
\switchcolumn
\begin{center}\textbf{\ref{sec:TheError}. The error}\end{center}
\switchcolumn

We now state the error in~\cite{Maniatis:2006fs} which was found during the process of
formalization. The authors of~\cite{Maniatis:2006fs} make the claim that
the stability condition expressed in Eq.~\ref{eq:K0_J2_J4} is equivalent to the following
condition, which we will call condition C:
~\\
\begin{thesisbox}
Condition C: For all vectors $\vec k$ satisfying $\|\vec k \|^ 2\leq 1$ we have
\begin{equation} \label{eq:incorrect_condition}
0 < J_4(\vec k) \text{ or } (J_4(\vec k) = 0 \text{ and } 0 \leq J_2(\vec k)).
\end{equation}
\end{thesisbox}

If true this would constitute a reduction in the complexity of
the stability condition to
\begin{align}
	\forall (k_1 \in \mathbb{R}), \forall (k_2 \in \mathbb{R}), \forall (k_3 \in \mathbb{R}), \_
\end{align}
This would be hugely beneficial as it lacks the existential quantifier $\exists$, making it
much easier to analyze.

Alas, this condition is not equivalent to the stability condition.
Because Theorem 1 of~\cite{Maniatis:2006fs} is a reworking of condition C, this also means that
Theorem 1 of~\cite{Maniatis:2006fs} is not correct (we do not show this formally as we believe it adds little benefit).
In particular, condition C is necessary for stability~\cref{code:necessary}, but
it is not sufficient.
\switchcolumn
\begin{derbox}{code:necessary}
\begin{codeLong}
lemma (*\physLeanLinks[quarticTermReduced\_nonneg\_of\_potentialIsStable]{TwoHiggsDoublet.quarticTermReduced\_nonneg\_of\_potentialIsStable}*)
    (P : PotentialParameters)
    (hP : PotentialIsStable P)
    (k : EuclideanSpace ℝ (Fin 3))
    (hk : ‖k‖ ^ 2 ≤ 1) :
    0 ≤ quarticTermReduced P k := by ...

lemma (*\physLeanLinks[massTermReduced\_pos\_of\_quarticTermReduced\_zero]{TwoHiggsDoublet.massTermReduced\_pos\_of\_quarticTermReduced\_zero\_potentialIsStable}*)
(*\physLeanLinks[\_potentialIsStable]{TwoHiggsDoublet.massTermReduced\_pos\_of\_quarticTermReduced\_zero\_potentialIsStable}*) (P : PotentialParameters)
    (hP : PotentialIsStable P)
    (k : EuclideanSpace ℝ (Fin 3))
    (hk : ‖k‖ ^ 2 ≤ 1)
    (hq : quarticTermReduced P k = 0) :
    0 ≤ massTermReduced P k := by ...
\end{codeLong}
\begin{derboxFoot}
Necessary conditions related to the stability of the potential.
\end{derboxFoot}
\end{derbox}
\switchcolumn

To prove it is not sufficient, we give a potential which is unstable but for which this
condition is satisfied. This potential is defined by the parameters~\cref{code:stabilityCounterExample}
\begin{align*}
&m_{11}^2 = m_{22}^2 = 0, \quad m_{12}^2 = i, \\
&\lambda_1 = \lambda_2 = \lambda_3 = \lambda_4 = \lambda_5 = 2,\\
&\lambda_6 = \lambda_7 = -2.
\end{align*}
\switchcolumn

%%%% Code snippet
\begin{inputbox}{code:stabilityCounterExample}
\begin{codeLong}
def (*\physLeanLinks[stabilityCounterExample]{TwoHiggsDoublet.PotentialParameters.stabilityCounterExample}*) : PotentialParameters :=
  {(0 : PotentialParameters) with
    m₁₂2 := Complex.I
    𝓵₁ := 2
    𝓵₂ := 2
    𝓵₃ := 2
    𝓵₄ := 2
    𝓵₅ := 2
    𝓵₆ := -2
    𝓵₇ := -2}
\end{codeLong}
\begin{inputboxFoot}
The parameters of a particular 2HDM potential.
\end{inputboxFoot}
\end{inputbox}
%%%% End of code snippet
\switchcolumn
After a bit of rearranging, this can be shown to correspond to the potential~\cref{code:potential_stabilityCounterExample}
\begin{equation}
V = 2\ \mathrm{Im} \Phi_1^\dagger \Phi_2 + ||\Phi_1 - \Phi_2||^4.
\end{equation}
\switchcolumn
\begin{derbox}{code:potential_stabilityCounterExample}
\begin{codeLong}
lemma (*\physLeanLinks[potential\_stabilityCounterExample]{TwoHiggsDoublet.potential\_stabilityCounterExample}*)
    (H : TwoHiggsDoublet) :
    potential .stabilityCounterExample H =
    2 * (⟪H.Φ1, H.Φ2⟫_ℂ).im +
    ‖H.Φ1 - H.Φ2‖ ^ 4 := by...
\end{codeLong}
\begin{derboxFoot}
The corresponding potential.
\end{derboxFoot}
\end{derbox}
\switchcolumn

Note that the quartic term of this potential is non-negative~\cref{code:quarticTerm_stabilityCounterExample_nonneg}, and when the quartic term is zero,
the quadratic term is also zero~\cref{code:massTerm_zero_of_quarticTerm_zero_stabilityCounterExample}. This is a potential we believe many readers
would naively think is stable.
\switchcolumn
\begin{derbox}{code:quarticTerm_stabilityCounterExample_nonneg}
\begin{codeLong}
lemma (*\physLeanLinks[quarticTerm\_stabilityCounterExample\_nonneg]{TwoHiggsDoublet.quarticTerm\_stabilityCounterExample\_nonneg}*)
    (H : TwoHiggsDoublet) : 0 ≤
    quarticTerm .stabilityCounterExample H := by...
\end{codeLong}
\begin{derboxFoot}
The quartic term is non-negative.
\end{derboxFoot}
\end{derbox}
\begin{derbox}{code:massTerm_zero_of_quarticTerm_zero_stabilityCounterExample}
\begin{codeLong}
lemma (*\physLeanLinks[massTerm\_zero\_of\_quarticTerm\_zero]{TwoHiggsDoublet.massTerm\_zero\_of\_quarticTerm\_zero\_stabilityCounterExample}*)
(*\physLeanLinks[\_stabilityCounterExample]{TwoHiggsDoublet.massTerm\_zero\_of\_quarticTerm\_zero\_stabilityCounterExample}*) (H : TwoHiggsDoublet)
    (h : quarticTerm
      .stabilityCounterExample H = 0) :
    massTerm .stabilityCounterExample H = 0 := by...
\end{codeLong}
\begin{derboxFoot}
The mass term is zero if the quartic term is zero.
\end{derboxFoot}
\end{derbox}
\switchcolumn

\switchcolumn
\begin{derbox}{code:stabilityCounterExample_not_potentialIsStable}
\begin{codeLong}
lemma (*\physLeanLinks[stabilityCounterExample\_not\_potentialIsStable]{TwoHiggsDoublet.stabilityCounterExample\_not\_potentialIsStable}*) :
    ¬ PotentialIsStable
      .stabilityCounterExample := by ...
\end{codeLong}
\begin{derboxFoot}
The potential is not stable. The proof is given in the text.
\end{derboxFoot}
\end{derbox}
\switchcolumn
This potential is in fact unstable~\cref{code:stabilityCounterExample_not_potentialIsStable},
as we will now prove.

To do this we must show that for all real numbers $c$ there exist Higgs doublets $\Phi_1, \Phi_2$ such that $V(\Phi_1, \Phi_2) < c$.
Thus, for $c$ a real number, let $t := \arctan(2 \sqrt{|c|+1})$ and define the Higgs doublets
\begin{equation}\label{eq:phi1_def}
\Phi_1 := \sqrt{\frac{\cos t}{4 t \sin^2 t}}\begin{pmatrix} 1\\ 0\end{pmatrix}
\end{equation}
and
\begin{equation}\label{eq:phi2_def}
\Phi_2 := \sqrt{\frac{\cos t}{4 t \sin^2 t}}\begin{pmatrix} 1-t\sin t+ i\, t \cos t \\ \sqrt{2 t\sin t - t^2}\end{pmatrix}.
\end{equation}
After a bit of (formalized) substitution and algebra, one can show that
\begin{equation}
V(\Phi_1, \Phi_2) = - 1/(4 * \tan^2 t) = - (|c| + 1) < c.
\end{equation}
This proves that the potential is unstable.

We now turn to show that condition C
holds for this potential. We remind the reader that Theorem 1 in~\cite{Maniatis:2006fs} is
a direct reworking of the statement that condition C is equivalent to the stability condition.

For this potential, we have that~\cref{code:massTermReduced_stabilityCounterExample}
\begin{equation}
J_2(\vec k) =  k_2,
\end{equation}
and~\cref{code:quarticTermReduced_stabilityCounterExample}
\begin{equation}
J_4(\vec k) = (1 - k_1)^2.
\end{equation}
\switchcolumn
\begin{derbox}{code:massTermReduced_stabilityCounterExample}
\begin{codeLong}
lemma (*\physLeanLinks[massTermReduced\_stabilityCounterExample]{TwoHiggsDoublet.massTermReduced\_stabilityCounterExample}*) (k : EuclideanSpace ℝ (Fin 3)) :
    massTermReduced .stabilityCounterExample k = k 1 := by...
\end{codeLong}
\begin{derboxFoot}
  The function $J_2$ for this potential.
\end{derboxFoot}
\end{derbox}
\begin{derbox}{code:quarticTermReduced_stabilityCounterExample}
\begin{codeLong}
lemma (*\physLeanLinks[quarticTermReduced\_stabilityCounterExample]{TwoHiggsDoublet.quarticTermReduced\_stabilityCounterExample}*) (k : EuclideanSpace ℝ (Fin 3)) :
    quarticTermReduced .stabilityCounterExample k = (1 - k 0) ^ 2 := by...
\end{codeLong}
\begin{derboxFoot}
  The function $J_4$ for this potential.
\end{derboxFoot}
\end{derbox}
\switchcolumn

From these expressions it is clear that for this potential that condition C is satisfied. In other words,
that for all $\vec k$ with $\||\vec k\|| \leq 1$,  $0 \leq J_4(\vec k)$.
and if $J_4(\vec k) = 0$  then $0 \leq J_2(\vec k)$. This latter point follows from the fact
that if $J_4(\vec k) = 0$
then $k_1 = 1$, and since $\|\vec k \| \leq 1$, this implies that $k_2 = 0$,
and thus $0 \leq J_2(\vec k)$. These points appear in the Lean code in the proof
of Snippet~\cref{code:forall_reduced_exists_not_potentialIsStable}. This is
the full statement that there exists a potential which is not stable but for which condition C is satisfied.
In full, there exists a potential which is not stable but for which
$0 \leq J_4(\vec k)$ and $(J_4(\vec k) = 0 \text{ implies } 0 \leq J_2(\vec k))$.
\switchcolumn
\begin{derbox}{code:forall_reduced_exists_not_potentialIsStable}
\begin{codeLong}
lemma (*\physLeanLinks[forall\_reduced\_exists\_not\_potentialIsStable]{TwoHiggsDoublet.forall\_reduced\_exists\_not\_potentialIsStable}*) : ∃ P, ¬ PotentialIsStable P ∧
    (∀ k : EuclideanSpace ℝ (Fin 3), ‖k‖ ^ 2 ≤ 1 →
    0 ≤ quarticTermReduced P k ∧ (quarticTermReduced P k = 0 →
    0 ≤ massTermReduced P k)) := by...
\end{codeLong}
\begin{derboxFoot}
There exists a potential which is not stable but for which
$0 \leq J_4(\vec k)$ and $(J_4(\vec k) = 0 \text{ implies } 0 \leq J_2(\vec k))$.
\end{derboxFoot}
\end{derbox}

\switchcolumn
In~\cite{Maniatis:2006fs}, they split the stability condition into a number
of difference cases: stable in the strong sense
$0 < J_4(\vec k)$ for all $k$; stable in the weak sense $0 \leq J_4(\vec k)$
for all $k$ and if $J_4(\vec k) = 0$ then $0 < J_2(\vec k)$; and at least stable in
the marginal sense corresponding to condition C above.
All these are contained in what we have called `stable' here,
meaning bounded from below. The error in~\cite{Maniatis:2006fs}
effects those potentials which are never strongly or weakly stable,
but which are marginally stable.

\section{Conclusion}

In this paper we hope to have convinced the reader
of one of the benefits of formalizing results from physics
into an ITP like Lean. That is, it can be helpful
in finding errors in the literature, and in giving confidence in physics results.

Naturally we have not formalized every result related to the 2HDM here, and
there is much future work to be done in this direction alone. For example,
next steps could include formalizing the various discrete group actions on the
2HDM potential, or formalizing what happens when one adds a Standard Model singlet in addition to the
extra Higgs doublet.
\end{paracol}
In addition to this, there are many formalizations to be done throughout high energy physics,
and physics more broadly. We encourage any reader interested in this to contribute to Physlib,
and join the community there.

Given the error in~\cite{Maniatis:2006fs} we now believe that the best
reduction in the complexity of the stability condition of the 2HDM potential is the one given in Eq.~\ref{eq:new_stability_condition}.
Further work in reducing this would be very interesting, and
it remains open as to whether this can be done.

There is a further point worth mentioning, which is that
the analysis above shows us that looking just at rays is not
always sufficient (particularly in the marginal case)
in determining whether a potential is stable or not.\footnote{We thank Igor Ivanov for
discussions on this point.}

\section*{Acknowledgments}
I thank the Physlib community and the broader Lean community for
endless useful conversations and help in the Lean Zulip.
I thank members of the Mathematical foundations of computation group
at the University of Bath for useful conversations on this paper,
in particular, to Ali Uncu. I thank Sven Krippendorf for discussions on Lean
and the subject of this paper. I thank the authors of~\cite{Maniatis:2006fs} and
Igor Ivanov for helpful discussions on this work.

\appendix
\section{Appendix: Tips on reading the Lean code}
\begin{enumerate}
\item Lean works on type theory, you can think of a type as similar to a set, although
  types are defined by specifying their elements. Some
  of the types in this paper are \myinline|TwoHiggsDoublet|, \myinline|HiggsVec|, \myinline|ℝ|,
  and \myinline|EuclideanSpace ℂ (Fin 2)|.
\item If \myinline|T| is a type, the notation \myinline|t : T| means that \myinline|t|
    is an element, or term, of the type \myinline|T|. An example seen throughout
    this paper is \myinline|H : TwoHiggsDoublet|.
\item A definition has the form \myinline|def name (input) : TypeOfDef := the_definition|.
\item A lemma has the form \myinline|lemma name (input) : statement := proof|. Here
   the proof is often done by tactics, however we do not show these here.
\item A \myinline|structure| like \myinline|TwoHiggsDoublet| has different fields,
   which can be accessed with dot notation. For example, if \myinline|H : TwoHiggsDoublet| then
   \myinline|H.Φ1| is the first Higgs doublet of \myinline|H|. This field notation
   extends to definitions in a given namespace (hidden here) but allows us to write
   \myinline|P.η| instead of \myinline|η P|.
\item A proposition in Lean is the same thing as a type, and a proof of a proposition is a term
  of the corresponding type. For example, if \myinline|P : PotentialParameters| then
  \myinline|PotentialIsStable P| is a proposition, and a proof of this proposition is an element of
  the type \myinline|PotentialIsStable P|.
\item The notation \myinline|→| can be read as "implies", or interpreted as a function type,
  in Lean these two interpretations are the same.
\item The notation \myinline|def name (a : A) : B := ...| is equivalent to
    \myinline|def name : A  → B|.
\item The notation \myinline|‖k‖| denotes the norm of the vector \myinline|k|, and
  the notation \myinline|⟪H.Φ1, H.Φ2⟫_ℂ| denotes the complex inner product of the two Higgs doublets.
\item For two types \myinline|T| and \myinline|S|, the notation \myinline|T ⊕ S| denotes the direct sum of the two types,
  which is a type whose elements are either an element of \myinline|T|, denoted \myinline|Sum.inl t| or \myinline|.inl t|,
   or an element of \myinline|S| denoted \myinline|Sum.inr s| or \myinline|.inr s|.
\end{enumerate}
\bibliographystyle{unsrturl}
\begin{spacing}{0.5}
\bibliography{MyBib}
\end{spacing}

%%%%%%%%%%%%%%
\end{document}